# Pros and cons of relativistic interstellar flight

Oleg G. Semyonov[1]

*State University of New York at Stony Brook, Stony Brook, NY 11794, USA*

Two technological problems must be solved before daring to interstellar flight: fuel and propulsion. The highest energy-density 'fuel' is antimatter in its solid or liquid state and this fuel is likely to be our primary choice for multi-ton relativistic rockets. High-energy ion thrusters powered by annihilation reactors promise superior performance in comparison with direct propulsion by annihilation products. However the power generator onboard can significantly increase the rocket dry mass thus limiting the achievable speed. Two physical factors that stand against our dream of the stars are thermodynamics and radiation hazard. Heat-disposing radiator also increases the rocket dry mass. Interstellar gas turns into oncoming flux of hard ionizing radiation at a relativistic speed of the rocket while the oncoming relativistic interstellar dust grains can cause mechanical damage. Economy and psychology will play a decisive role in voting for or against the manned interstellar flights.
*Keywords; space, interstellar, relativistic rocket, antimatter*

## I. Introduction

This paper is inspired by an enthusiastic article by D Foley and W. S. Weed published in Discover Magazine, August 01, 2003. The authors advocate the interstellar travel is an enterprise which is relatively "not hard" to perform. In fact, the future of star journeys is not so rosy. The detailed discussion of technical and physical issues regarding the relativistic interstellar flight can be found in [1]. Here we consider rockets with energy source and propellant on board [1] putting aside the concepts of externally powered spacecrafts and discuss shortly the challenges we will inevitably meet on our road to the stars as well as the physical factors that can limit or even prevent our expansion beyond the solar system. The future of relativistic interstellar voyaging depends on technology, economy, and social psychology. Ordinary rocket fuel (chemical, nuclear, and even thermonuclear) is not able to accelerate a multi-ton rocket to a relativistic speed above $0.1c$, where $c$ is the speed of light, because only a tiny fraction of mass of these fuels turns into propulsion kinetic energy and a copious mass-rate efflux is needed to get a necessary thrust, which limits the achievable speed of a rocket. A hundred-ton rocket equipped with the ion thruster and powered by a nuclear reactor can reach the speed of 200 to 300 km/s in several months, which is acceptable for establishing interplanetary transportation within our solar system. [1] To produce a sufficient thrust for many years of space flight and to accelerate a multi-ton rocket to a relativistic speed we need an ultimate fuel (Table 1) – antimatter. Virtually all mass converts into energy in its annihilation with ordinary matter to be used for propulsion. Many technical problems are to be solved before we risk flying beyond the solar system and among them antimatter production, safe storage, and implementation for propulsion are among the most challenging.

---
[1] Research Professor (retired), Ph. D. oleg.semyonov.1@stonybrook.edu

**Table 1 Energy density of commonly used and potentially usable fuels.**

| Fuel | Energy density, MJ/kg |
|---|---|
| Antimatter | 89,900,000.000 |
| Hydrogen fusion | 650,000,000 |
| DT fusion | 340,000,000 |
| U-235 | 88,000,000 |
| Natural U | 81,000,000 |
| Pu-238 | 15,000,000 |
| H chemical burning | 140 |
| Rocket chemical fuels | 50 |
| Gasoline | 44 |
| Coal | 24 |
| Wood | 16 |

Finance is a driving force of any enterprise. The great geographic discoveries in the middle Ages would be impossible without investments in ships building and supplies acquisition for the long journey to the unknown land. These investments, no matter how risky, were given by kings and reach men to the dreamers and adventurers in anticipation of a profit from the newly discovered lands. Investments are evenly important in our cosmic epoch especially for the costly deep-space expeditions. Psychologically, people are in favor of space exploration. Our curiosity and aspiration to new discoveries play a major role in planning the space missions in addition to possible profitability and sometimes even contrary to the latter due to political reasons.

## 2. Pros: Technology advance and social spirit

Two concepts of direct rocket propulsion by the products of matter-antimatter annihilation have been proposed: 1) photon rocket propelled by a beam of gamma-photons emitted in the process of electron-positron annihilation and reflected by a mirror [2] and 2) meson rocket propelled by a jet of charged π and μ-mesons produced by annihilating protons and antiprotons inside a magnetic nozzle. [3] Both concepts are hardly realizable technically and do not promise high propulsion efficiency due to relatively small cross-sections of electron-positron and proton-antiproton annihilation resulting in lengthy "annihilation zone" and, as a consequence, in poor alignment of the propulsion jet. [1, 4] Propulsion by a stream of high-energy ions [1, 4] promises much better performance regarding the efficiency of annihilation energy conversion into propulsion energy and nearly ideal alignment of the propulsion jet of ions. The method is based on existing technology of linear ion accelerators which generate almost parallel beams of high-energy ions. Transition to relativistic ion propulsion is beneficial because it leads to a significant reduction in propellant mass flow resulting in huge economy of propellant and launching mass of the rocket. The thrust produced by the relativistic ions $F_i = N\gamma_i m_i \beta_i c$, where $N$ is the number of exhausted ions per second, $m_i$ is the rest mass of an ion, $\beta_i = v_i/c$ is the Einstein velocity factor of the ions in the rocket coordinate frame, $v_i$ is the velocity of exhausted ions in the rocket coordinate frame, and $\gamma_i = 1/(1 - \beta_i^2)^{1/2}$. It increases proportionally to $\gamma_i \beta_i$ allowing lower mass exhaust thus propellant economy. A trade between kinetic power of the ion jet and achievable rocket velocity allows choosing an optimal exhaust ion velocity $v_i$. Low-energy ion propulsion is currently under development by NASA and the ion thrusters are routinely used on satellites and interplanetary modules for stationkeeping/maneuvering as well as for interplanetary probes propulsion. [5, 6] Apparently, our first step to deep-space will be boosting this technology to a higher exhaust ions velocity and higher propulsion power implementing the small nuclear reactors as power generators to prove feasibility of the high-energy ion thrusters. It is still a non-relativistic case and the rocket speed $\beta$ can be determined from the well-known equation:

$$\frac{M}{M_0} = \exp\left(-\frac{\beta}{\beta_i \gamma_1}\right), \tag{1}$$

where $M_0$ is the rocket launching mass, $M$ is its instant (leftover) mass, $\beta_i = v_i/c$ is the exhaust velocity factor of ions jettisoned by the ion thruster, and $\gamma_i = 1/(1 - \beta_i^2)^{1/2}$. At the moment, when a half of the launching mass is exhausted to propulsion ($M/M_o = (M_0 - \dot{M}\tau)/M_0 = 0.5$), the achievable velocity $\beta_{0.5} = v_{0.5}/c$ of a hundred-ton rocket powered by a 600-megawatt small nuclear reactor with the conversion efficiency $\varepsilon = 50\%$ to kinetic power $P_k = \varepsilon P$ of the ion propulsion jet is shown in Fig. 1a as a function of $\beta_i$. The time of flight $\tau$ to the moment when the rocket exhausts a half of its launching mass is plotted in Fig 1b as a function of $\beta_i$ for the rocket launching masses of 100, 200, and 300 tons on assumption of constant thrust. One can see that a higher velocity of jet ions allows reaching a higher rocket velocity at the moment when a half of the rocket mass is exhausted but the price is a longer acceleration time. A trade between the achievable rocket velocity and the tolerable acceleration time will determine the optimal ion exhaust velocity and the amount of propellant saved for the rocket braking to cancel its speed upon arriving to the destination. [1]

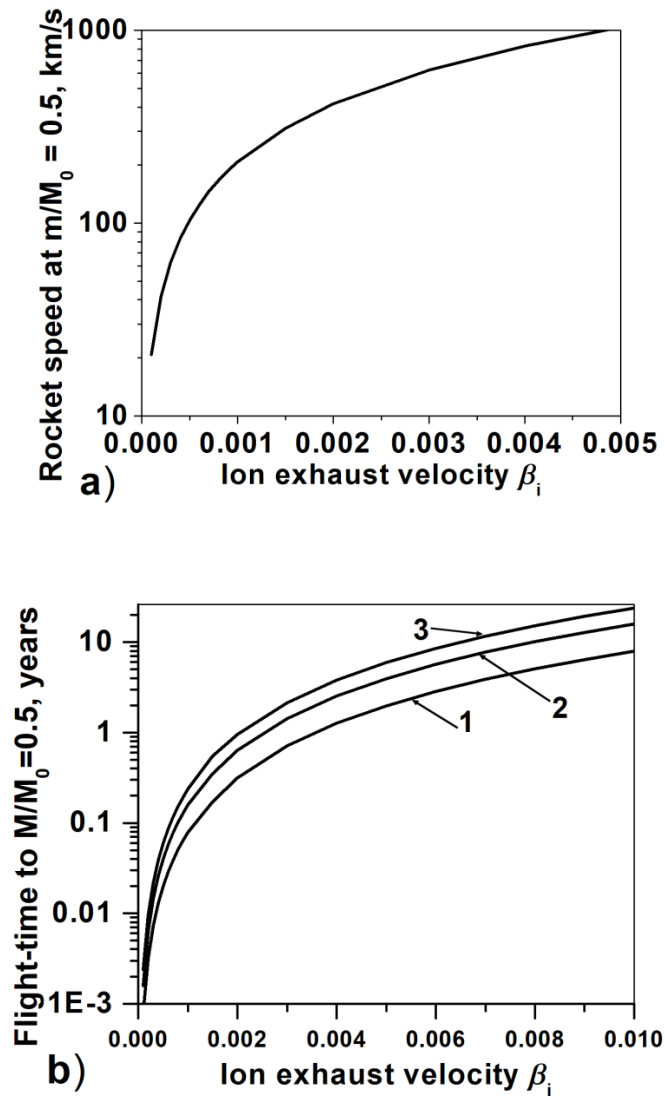

Fig.1 a) Achievable velocity of an interplanetary rocket ($M_0$ = 100 tons) propelled by a jet of accelerated protons and powered by a 600-MW small nuclear reactor with the conversion efficiency $\varepsilon$ = 50%, when a half of rocket launching mass is exhausted for propulsion, as a function of the proton exhaust velocity $\beta_i$ in a range between $\beta_i$ = 0.0001 and 0.005
b) Rocket time of flight (years) to the point when it exhausts a half of its launching mass to propulsion ($M/M_0$ = 0.5) as a function of proton exhaust velocity. 1 – $M_0$ = 100 tons, 2 – $M_0$ = 200 tons, and 3 – $M_0$ = 300 tons.

Owing to its highest energy density, antimatter in its condensed state (liquid or solid) is the most relevant fuel for powering a multi-ton relativistic rocket and the main candidate is antihydrogen. Obviously, annihilation reactor cannot be designed without solving the problem of antimatter production and storage. Commonly, antiprotons (nuclei of hydrogen atoms) are generated upon irradiation of metallic targets by high-energy protons. [7] The obtained antiprotons are slowed down and mixed with positrons from a separate source for mutual recombination into antihydrogen atoms. Efficiency of the process is very poor, so the production rate of antihydrogen atoms must be increased by many orders of magnitude. The tightest bottleneck so far is the rate of antihytdrogen atoms production in the mixed beams of antiprotons and positrons. According to calculations, the production of antihydrogen atoms can be increased by several orders of magnitude, if antiprotons are passing through gas of positronium each atom of which consists of an electron and a positron). [8] Next, atomic antihydrogen should be prompted to recombine to molecular antihydrogen and then stimulated to condense into liquid or solid antihydrogen at a cryogenic temperature. Since liquid or solid antihydrogen in analogy with hydrogen consists mostly of diamagnetic antiparahydrogen, its inherent diamagnetism can be used for storage in a closed configuration formed by a gradient magnetic field. A magnetic barrier for diamagnetic medium can be induced by an array of high-temperature superconductor coils with alternating direction of current assembled near the inner wall of a tank. [1, 9] The same approach can also be implemented in a hose for antihydrogen delivery to the annihilation reactor.

The relativistic rocket equation, which accounts for the loss of mass of antihydrogen fuel and propellant, is: [1, 4]

$$\frac{M}{M_0} = \left(\frac{1-\beta}{1+\beta}\right)^{\frac{(1+(\chi_i-1)/\varepsilon)}{2\gamma_i\beta_i}} . \qquad (2)$$

Transforming the equation (2),

$$\beta = \frac{1-\left(\dfrac{M}{M_0}\right)^{\eta}}{1+\left(\dfrac{M}{M_0}\right)^{\eta}}, \text{ where } \eta = \frac{2\gamma_i \beta_i}{1+\dfrac{\gamma_i-1}{\varepsilon}}, \tag{3}$$

where, $M$ is the instant mass of the rocket, $M_0$ is its launching mass, $\beta = v/c$ is the rocket map-velocity factor, $\beta_i$ is the relativistic velocity factor of efflux ions in the rocket coordinate frame, $\gamma_b = (1 - \beta_i^2)^{-1/2}$, and $\varepsilon$ is the conversion efficiency (ratio of the efflux jet kinetic power to the power of hydrogen-antihydrogen annihilation reactor). Assuming a constant thrust during the acceleration stage, so that $\dot{M} = \dot{M}_{ann} + \dot{M}_{jet} =$ Const. and $M = M_0 - \dot{M}\tau$, where $\tau$ is the proper time of flight by the rocket clock, $P_{jet} = \varepsilon P$ is the ion jet kinetic power, $P$ is the reactor power, and $\varepsilon$ is the reactor power conversion efficiency to the propulsion jet kinetic power, the time of flight to the moment, when the mass ratio $M/M_0 = a$ for a chosen exhaust ions velocity, and the achieved rocket speed at this moment $\beta_a$ accounting for the mass loss due to annihilation and for propulsion can be expressed as

$$\tau_a = \frac{M_0 c^2 (\gamma_i - 1)}{P(\gamma_i - 1 + \varepsilon)}[1 - \left(\frac{1-\beta_a}{1+\beta_a}\right)^{\frac{(1+(\chi_i-1)/\varepsilon)}{2\gamma_i \beta_i}}]. \tag{4}$$

The achievable rocket map-velocity at the moments when $M/M_0 = 0.5$ and $M/M_0 = 0.25$ is plotted in Fig. 2a as a function of proper velocity of efflux jet ions $\beta_i$. The proper time of flight until ether $M/M_0 = 0.5$ or $M/M_0 = 0.25$ is reached is plotted in Fig. 2b and 2c as a function of proper velocity of the efflux ions in the rocket coordinate frame for two launching masses of 1000 and 10000 tons and for the reactor power of $10^{12}$ watts (one terawatt), $10^{13}$ watts, and $10^{14}$ watts. The corresponding graphs of the rocket velocity as a function of proper time of flight and the distance covered during the acceleration stage can be found in [1, 4].

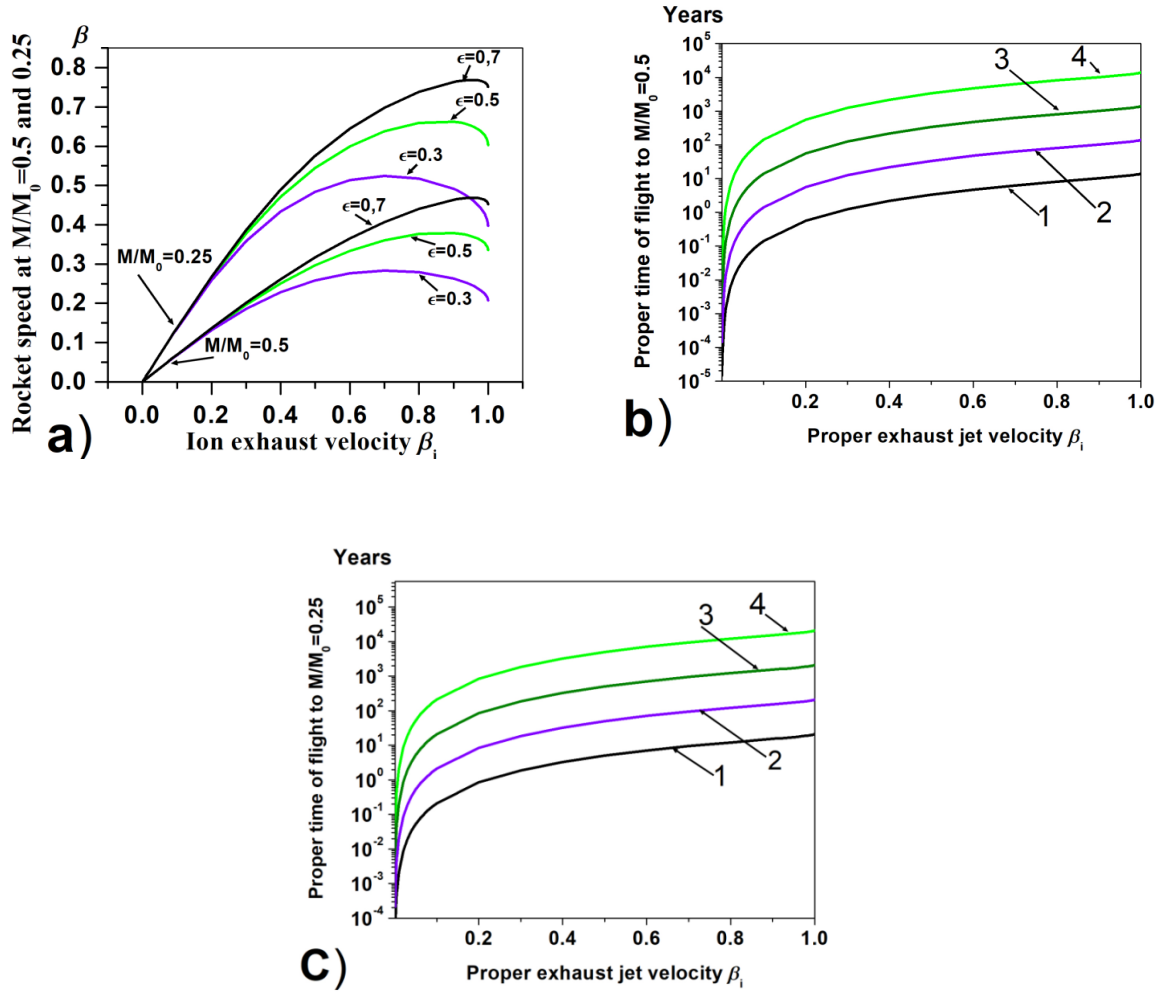

Fig. 2 a) Rocket speed at the moments when either a half or three quarters of the rocket launching mass is exhausted for propulsion (M/M₀ = 0.5 and M/M₀ = 0.25) as a function of proper velocity of the ion jet for the values of propulsion efficiency ε = 0.3, 0.5, and 0.7.
b) Time of flight to the moment, when a half of the rocket launching mass is exhausted, as a function of proper efflux velocity: 1 – $M_0$ = 1000 tons and $P$ = 100 terawatts; 2 – $M_0$ = 1000 tons and $P$ = 10 terawatts or $M_0$ = 10000 tons and $P$ = 100 terawatts; 3 – $M_0$ = 1000 tons and $P$ = 1 terawatt or $M_0$ = 10000 tons and $P$ = 10 terawatts; 4 – $M_0$ = 10000 tons and $P$ = 1 terawatt. Propulsion efficiency ε = 0.5.
c) Time of flight to the moment when three quarters of the rocket launching mass is spent for propulsion (M = 0,25M₀) as a function of exhaust jet proper velocity: 1 – $M_0$ = 1000 tons and $P$ = 100 terawatts; 2 – $M_0$ = 1000 tons and $P$ = 10 terawatts or $M_0$ = 10000 tons and $P$ = 100 terawatts; 3 – $M_0$ = 1000 tons and $P$ = 1 terawatt or $M_0$ = 10000 tons and $P$ = 10 terawatts; 4 – $M_0$ = 10000 tons and $P$ = 1 terawatt. Propulsion efficiency ε = 0.5.

The graphs demonstrate principal potentiality of reaching the nearby stars within the local low-density cavity [10] in a reasonably moderate proper time of flight from years to tens of years, if the launching mass of a rocket $M_0$ is below 10000 tons, its dry mass is less than 0.25 $M_0$, and the propulsion engine power is well above 1 terawatt. The anticipated technical challenges include: design of a multi-terawatt propulsion engine squeezed into a limited volume and having its mass below the allowable dry mass of the rocket, search for a method to produce and store the abandon amount of antihydrogen, and development of a multi-terawatt propulsion engine that uses antimatter fuel energy for thrust production. Also, the efficient method to dispose the waste heat from the engine has to be found in order to avoid rocket overheating (see the next section)

Psychologically, people are in favor of studying other worlds, at least an educated part of humanity, and many developed countries can support the space missions financially within their budget. Among the anticipated benefits from space exploration, science and engineering are foreground. Understanding how stars with planets appeared and deciphering the origin of our universe evolutionally predisposed for giving birth to biological life, which, in its turn, gives birth to mind with its creative ability to reshape and reorganize matter, will be revolutionary for the future cosmic civilization. Colonization of other planets and the long anticipated encounter with aliens could follow. Probably, colonization of other worlds will even be a priority in our expansion first to the planets of solar system and then beyond it to guarantee survival of human species. We have may be about 1000 to 10000 years for salvation of human species and "by that time we should have spread out into space, and to other stars, so a disaster on Earth would not mean the end of the human race," as Stephen Hawking told BBC. [11]

It is worth noting that the ancient people have been once in a similar challenging situation. Marine expeditions in the middle Ages made a surprising discovery: almost every island on our planet was found inhabited, even as remote as Hawaii, Fiji, Polynesia, Ester Island, and New Zealand. How could it happen? All evidences show that human species appeared in Africa and expanded from there to Europe, Asia, and even to America on dry land, which is explainable. However, the islands in the middle of the oceans have never been connected with the mainland at least since the appearance of human species in Africa. Putting aside an activity of benevolent aliens who developed humans by deliberate selection in the African zoo and then distributed them over the globe to guarantee their survival in possible local catastrophes, we have to accept the fact that a civilization of prehistoric see-farers existed may be before the emergence of the Sumerians and the ancient Egyptians. The marine expeditions undertaken by those 'primitive' tribes far away from their homelands, who sailed or rowed their boats into the oceans beyond the vastness of their eucumene (known universe), are amazing because it is virtually impossible to imagine that they possessed a navigation map and have knowledge of coordinates of all islands and continents over the globe not mentioning trade winds and ocean currents. Each expedition should consist of hundreds of men and women otherwise their survival in the new land would be impossible. Imagine the amount of food and water they needed to take with them. How many died in stormy oceans and how many disappeared without a trace? What made them so willingly embark on a voyage in search of new lands amid the oceans?

Economical arguments on the pro side of the dilemma of interstellar journeys are the same as they were in the ancient times. Aside of searching for a safe harbor in an unfortunate case of doomsday on our planet, colonization of life-supporting planets can lead to a technological boom and huge advance in science. If we meet the benevolent intelligent species who are much more advanced technologically, the benefits can be invaluable but the dangers are unpredictable.

### 3. Cons: Physical and social factors

*3.1 Heat*

Thermodynamics dictates disposal of excess heat from every system in which a thermodynamic process occurs. According to the concept of direct propulsion by annihilation products, heat accumulates on the mirror of a photon rocket due to absorption of photons or on the magnetic nozzle frame and protective shield of a meson rocket due to absorption of gamma-radiation from the decaying $\pi^0$-mesons. In the concept of relativistic ion propulsion, thermal energy is accumulated in the reactor/turbine walls and carried with the exhaust gas after the turbine. The only way to dispose heat in space is by thermal radiation. The surface area $S_r$ of a radiator and its mass $M_r$ can be estimated from:

$$M_r = \rho S_r \delta = \frac{\rho \delta P_t (1 - \varepsilon)}{\eta \sigma T^4}, \tag{5}$$

where $\rho$ is the density of the radiator material, $\delta$ is the thickness of the radiator walls, $P_t$ is the total power released in the process of fuel burning, $\varepsilon = P_{exh}/P_t$ is the propulsion efficiency, $P_{exh}$ is the kinetic power of the propulsion jet, $\eta$ is the emissivity of the radiator material, $\sigma$ is the Stephan-Boltzmann constant, $T$ is the radiator temperature, and $\eta \sigma T^4$ is the thermal radiation power from the unit surface of a radiator. Both $S_r$ and $M_r$ are proportional to reactor power for a given ε and sharply diminish with the increasing temperature of the radiator. Propulsion efficiency ε of a meson rocket accounts for energy loss to gamma radiation, pions seeping through the bottleneck of the magnetic nozzle, and their poor alignment in the exhaust jet. [1] In the case of high-energy ion propulsion powered by a reactor and turbines, the thrust engine can be treated as a heat machine. Its efficiency cannot exceed a thermodynamic limit $\varepsilon_t = (T_0 - T)/T_0$, where $T_0$ is the temperature of gas delivered from the reactor to the turbines and $T$ is the temperature of gas entering the radiator. Here we have a dilemma. To boost the propulsion efficiency, $T_0$ must be as high as possible whilst $T$ must be as low as possible. However: 1) $T_0$ is limited by the melting point of the reactor material and by the thermal resistance of the turbine blades; 2) thermal emission per unit surface of a radiator is proportional to $T^4$ thus a higher radiator temperature is beneficial for a higher emission power per unit area thus allowing reduction of total area and mass of radiator. The last condition contradicts to our aspiration for higher efficiency ε so a compromise has to be found between the propulsion efficiency and the tolerable radiator mass. Radiator in the form of a lace of thin-walled tubes would allow increasing its radiating surface and emission power. It would be technically simple to assemble but it should be kept in mind that its mass grows rapidly with the emission area proportional to the engine power. The radiator can be unacceptably large and massive for the engine powerl above one terawatt. [1, 4]

*3.2 Ionizing radiation*

Space beyond the earth's atmosphere is not just an empty void. Interstellar space in every galaxy including our Milky Way is filled with rarefied gas and dust even in the low-density regions such as a local low-density cavity about 400 light-years in size around our solar system. [10] Concentration of neutral and ionized atoms and molecules (mostly hydrogen and helium) in the local cavity is ~0.3 cm$^{-3}$ ($3 \times 10^5$ m$^{-3}$) and concentration of dust grains of $10^{-5} \div 10^{-6}$ m in size with their masses ranged from $10^{-17}$ to $10^{-20}$ kg is from $10^{-8}$ m$^{-3}$ in the low-density regions to $10^{-5}$ m$^{-3}$ in the dense clouds of the galactic arms. [13] When a rocket accelerates to a velocity $v$ relative to a map-frame (reference coordinate frame fixed to a point in space, which can be considered at rest), all gaseous components and dust grains start moving relative to the rocket with the same velocity $v$ in opposite direction forming a headwind just like one can experience in a moving car. When the rocket accelerates to a relativistic speed closer to the speed of light, the headwind is relativistic, too. As a result, a new physical phenomenon appears: otherwise innocuous and rare interstellar gas becomes an ongoing stream of high-energy ions, atoms and dust granules. Their kinetic energy in the rocket coordinate frame $E_k = m_0c^2(\gamma - 1)$, where $m_0$ is the mass of rest of a corresponding gas particle or dust grain, $c$ is the speed of light in vacuum, $\gamma = (1 - \beta^2)^{-1/2}$, and $\beta = v/c$, increases with $v \to c$. Kinetic energy of ionized and neutral atoms of hydrogen ~100 MeV or more, when $v > 0.5c$, which is characteristic of hard ionizing radiation. Despite rarefaction of interstellar gas, the flux of relativistic electrons, ions and molecules in the rocket frame $N = \gamma n v$ exceeds $10^9$ per square centimeter per second ($10^{15}$ m$^{-2}$s$^{-1}$) at the rocket velocity $v \geq 0.3c$ and the rate of radiation absorption dose in the tissue of an unprotected astronaut is ~ $10^4$ rems per second, which is comparable to that in the core of a nuclear reactors. [1, 12] Relativistic factor $\gamma$ in the expression for $N$ is due to relativistic contraction of longitudinal space scale along the rocket $v$-vector resulting in corresponding one-dimensional increase in proper concentration $n' = \gamma n$ with respect to map-concentration.[2] The flux of atomic particles and the dose rate without radiation protection are plotted in Fig. 3 as functions of rocket map-velocity factor $\beta$.

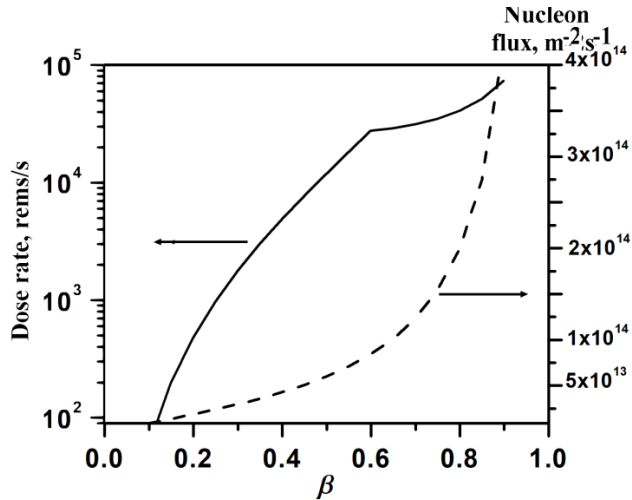

Fig.3 Flux of the interstellar atoms and ions per square meter per second (dashed) and radiation dose rate (rems per second) obtained by an unprotected astronaut as functions of rocket map-velocity $\beta$. The brake on the dose rate graph near $\beta \sim 0.6$ is corresponds to the velocity at which the penetration depth of nucleons starts exceeding the thickness of human torso.

---

[2] It can be understood from the following. $N_0 = nL = nvt$ in the map-frame and $N_0 = n'dL' = n'v\tau$ in the rocket coordinate frame, where $N_0$ is the total number of gas particles per square meter along the whole distance $L$ to the destination star. $N_0$ is the same in both coordinate frames.

The official safe radiation dose is 5 rems according to NIST safety regulations. The dose $10^2$ rem is considered dangerous due to high probability to develop cancer, and the dose $10^3$ rem is hundred percents lethal. According to the graph in Fig. 3, the lethal dose will be accumulated in the astronaut body in a fraction of a second, if $v \geq 0.3c$. To reduce the dose rate, a radiation-absorbing shield has to be mounted in front of the rocket inevitably adding to the rocket dry mass in addition to the thermal radiator. A relatively light-weight shield comprising a magnetic system and an electron stripper [1, 12] can protect the rocket from this relativistic flux of ionized and neutral components of interstellar gas. It is not clear however, if protection against relativistic dust granules is possible at all without a thick and massive bulge of solid material in front of the rocket. The oncoming dust grains will bombard the frontal parts of the rocket with a rate exceeding 100 per square meter per second, if $\beta > 0.3$. They can pierce through the protective shield and damage the rocket body, despite their smallness. The impact of relativistic multi-atomic grains on the materials has never been studied because we do not possess a means for accelerating the multi-atomic granules to relativistic velocities. Possibly, a relatively thin shell of constantly renewable material such as a layer of freezing ice permanently grown on a mesh of thin tubes with refrigerating liquid can compensate the loss of material due to dust bombardment, serving simultaneously an electron stripper for neutral atoms in the oncoming relativistic gas flow.

Taking into account the physical factors above, we can sketch an interstellar relativistic rocket comprising the following necessary elements: 1- high-energy ion thruster made of linear accelerators of high-energy ions), 2 - propellant tank and a container for the medium annihilating with antihydrogen (propellant can be used for annihilation, too), 3 - refrigerators to maintain a cryogenic temperature of antihydrogen in tanks and a frontal shield. 4 - thermal insulators, 5 – electrical generator, 6 - annihilation reactor to power the propulsion accelerators and auxiliary equipment , 7) control bridge with or without crew, 8 – crew quarters or AI module, 9 – frontal protective shield against ionizing radiation, 10 – antidust shielding system, 11 - heat radiator, 12 – antihydrogen tanks. A conceptual interstellar relativistic rocket is shown in Fig. 4. Two radiators are shown but it could be four or more of them to increase their total radiating surface. Hence the radiators must also be protected against the incoming relativistic gas and dust, the additional shields has to be mounted in front of them (shown as relatively thin radial protruding from the main shield assembly).

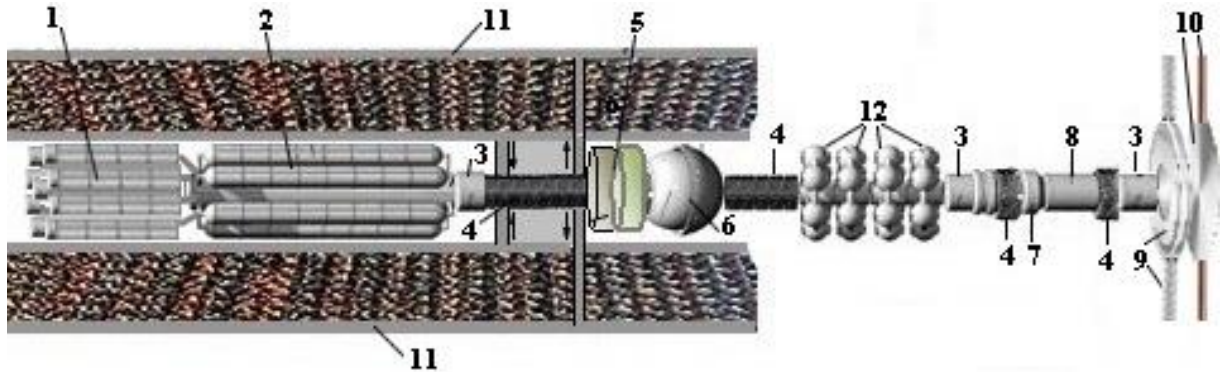

Fig. 4 Conceptual relativistic rocket: 1 – ion accelerators, 2 – propellant tanks, 3 – refrigerators, 4 – heat insulators, 5 - turbines and power generator, 6 – annihilation reactor, 7 – bridge, 8 – crew quarters or robotic module, 9 – magnetic shield, 10 – electron stripper, 11 – heat radiator, 12 – antihydrogen tanks. The sketch contains some basic elements borrowed from the article by D. Foley in Discover Magazine, 2003: http://discovermagazine.com/2003/aug/cover .

The oncoming flux of relativistic gas and dust makes dangerous any maneuver of the rocket cruising with a relativistic speed. If rocket comes out of protective 'shadow' of the shield during a maneuver, it will be exposed to all fury of the oncoming radiation flux. [1, 14] The same is also true for a rocket performing a turnaround maneuver to start braking which will require reversal of the exhaust ion jet. Simple rotation of a rocket by $180°$ will not work because it means losing a protective shadow of the shield during upside-down rotation and even after the rocket turnaround because the protective shield cannot be placed in the exhaust jet directed forward now. Estimations showed that the braking jet itself is unable to sweep completely interstellar gas and dust out of the rocket way [1] thus the parts of the propulsion thruster will be exposed to relativistic gas and dust bombardment. A possible solution is to rotate independently the aft accelerators in order to redirect the ion exhaust jets and make them flow mainly ahead but at an angle relative to the rocket axis of symmetry in order to produce the braking momentum avoiding any damage of rocket construction elements including the protective shield (Fig. 5). This way, the whole rocket can be still hidden from the gas and dust flow behind the frontal shield.

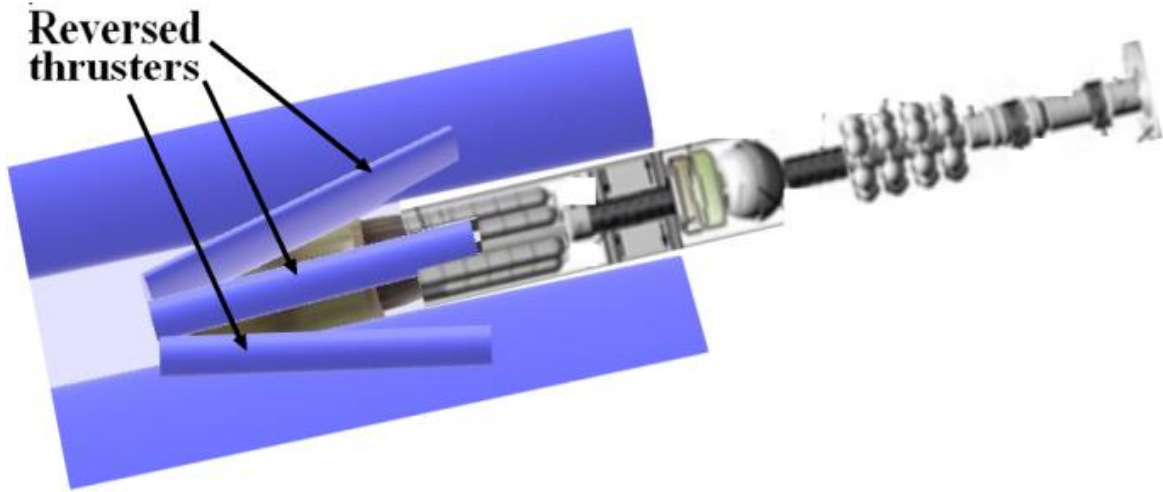

Fig. 5 The rocket at its braking stage with the thrusters (ion accelerators) reversed.

*3.3 Flight duration*

According to the theory of relativity, the rocket map-speed can never exceed the speed of light thus the flight-time by the Earth's clock in years cannot be smaller than the distance to the destination star in light-years. Apparently, a practically achievable speed will be below $0.7c$ because of limitations on the propulsion power for a given fuel capacity (Fig. 2) thus the relativistic time contraction factor γ will not exceed γ = 1.4 and any hope for a significant relativistic shortening of the time of flight by the rocket clock is virtually futile. For example, the recently discovered planetary system [15] with the Earth-like planets located at the distance of 39 light-years can be reached by a rocket, cruising with the speed of $0.7c$, in 55 years plus another five to ten years for acceleration and deceleration. By rocket clock, the total time of flight will be about 40 years. Add the return flight, and we get a mission which will take a whole human life to accomplish. Will a crew be psychologically stable spending all their life in a compact habitat? Will anabiosis, if possible in principle, be safe and acceptable? Imagine, how much water, food, and other supplies will be needed for a mission. Imagine now the cost of a mission. Here, economics and psychology will obviously be among the most decisive factors to voting against the manned flights.

Another consequence of huge distances between stars and galaxies is that we see other star systems in their distant past because all the photons and other carriers of information, we detect now, were emitted from them many years ago. Correspondingly, they see us (solar system and Earth) in our past because the information-carrying particles moving at the speed of light need time to reach them, and this time is equal, if measured in our years, to the distance between us and them in light-years. We are separated from aliens not only by huge distances but also by a time abyss. May be, here is the answer to the conundrum of Great Silence of the universe. [16] We do not detect signals from them because either they have not been sufficiently advanced yet in their past, we watch now, or they had been technologically advanced but did not notice us yet (our radio, laser, and other technological activity) in their past to become aware of us and be interested in sending us a signal because they saw (and see now still) our distant past, when no technological activity on our planet existed.

## 4. Conclusion: Do we have a hope?

Possibly, our only hope to reach other stars is robotic rockets. Artificial intelligence (AI) is actively discussed now in science and engineering and its indispensability for interstellar missions has been passionately promoted by Keith B. Wiley [17]. No duration of mission is a problem and no return flight is needed because all information can be sent back to Earth by radio or optical sharp-beam transmitters. No life supporting supplies and equipment are needed and no psychological problems will arise in AI, hopefully. It will be a huge economy in rocket mass and cost of the mission. The parts of rocket which cannot be slashed out are protective shield against nucleonic radiation and dust and thermal radiator for excess heat disposal.

Most likely, robotic flights will be undertaken mostly within the local low density cavity [10] because flying with a relativistic speed through the dense clouds of gas and dust in the galactic arms can be catastrophic. Overgalaxy flights above the galactic plane would take many thousands years [14] It is difficult to imagine, if someone will be interested in financing the missions requiring many hundreds or thousands of years to accomplish. We can hardly count on a physical encounter with aliens in a foreseen future unless an advanced alien civilization is our neighbor in the local cavity. May be their absence in solar neighborhood is our good luck because a physical contact can be disastrous for both species and not due to mutual hostility but because each species carries their own microbes and viruses, so an encounter will expose both counterparties to the whole biohazard of their planets with no immunity against alien micro-life. The consequences were predicted more than a century ago by H. G. Wells in his sci-fi novel [18] reincarnated in a 2005 science-fiction movie. Even primitive life on another planet can be poisonous for humans, and visiting biologically active planets is not like boarding a cruise liner to a banana republic. Even sending an unmanned probe to such a planet and then returning it back to Earth could be disastrous.

We considered rockets with energy source and propellant on board [1] putting aside the concepts of externally powered spacecrafts. Among other ideas of paving the road to the stars, the concept of a one-gram probe carrying a wafer for data processing and propelled by a mirror (optical sail) reflecting a laser beam from a stationary platform can be mentioned. [19] Without going into details of optics and mechanics, the following weak points of the program can be mentioned: 1) probe is unable to cancel its speed upon arrival to the destination thus a flyby missions at relativistic velocity can only be performed, 2) data transmission from a gram-ranged probe back to Earth through huge distances is not an easy task, and 3) probe moving with relativistic speed will be bombarded by oncoming relativistic flow of atoms of interstellar gas and dust particles residing in interstellar space which becomes a flux of ionizing radiation of extremely high intensity (see Section 2.2 of the manuscript) causing degradation of wafers and other electronics in hours. [1, 12] A probe must be protected by a frontal shield against hard radiation and relativistic dust and such a shield can weigh may be hundreds of grams or more to survive years of sputtering by ionic radiation and dust bombardment.

To accelerate a multi-ton rocket to a relativistic speed, the needed propulsion power is huge (terawatts or tens of terawatts). The engine mass (reactor with turbines plus ion thruster) should not desirably exceed, say, a half of the rocket dry mass leaving some room for heat radiator, frontal protective shield, and rocket body with needed equipment. It is a challenging task and it is not clear if this goal is achievable. For comparison, electrical power generators currently in use at nuclear power plants have the capacity below ten gigawatts [20] and small and medium nuclear reactors onboard of submarines and airplane carriers are below one gigawatt. [21] Submarine small reactors are quite compact and relatively lightweight (~ 10 - 30 tons). [21, 22] Gigavatt-range annihilation reactor can be even lighter because it does not contain nuclear fuel inside, does not produce neutrons thus not need for heavyweight shielding of neutron radiation (may be gamma-absorbing shield will be needed to protect the rocket bridge), no radioactive waste, no cooling water (liquid light metal, molten salts or gas as heat carriers can be implemented). However, it is difficult at the moment to scale its mass as a function of power to terawatt range and above. Rocket launching masses from 1000 to 10000 tons chosen above allow reaching velocities between $0.3c$ and $0.7c$ in a reasonable time from years to tens of years (Fig.

2) with a ten-to-hundred terawatts annihilation-propulsion engine and can be considered as our goal. Future will show if this goal is achievable. Otherwise, the missions even to nearby stars will take hundreds to thousands of years – the doomsday scenario.

It is worth mentioning also, that the physical laws are the same across the universe and equally true for us and for aliens no matter what they are or where they live. They meet the same problems in their space-faring activity, if they are able to undertake interstellar flights, and this is a blow to the idea of their frequent visits to Earth. If the advanced civilization existed in the local low-density cavity in the past, which is hardly possible otherwise we would notice their radio activity now, their visits to Earth couldn't be done more often than once in hundred or several hundreds of years. If they dwell somewhere in our galaxy, the time span between their possible visits cannot be less than tens of thousands or even hundreds of thousands of years in the best scenario. If no intelligent life exists in our galaxy except us and the advanced civilizations are somewhere in other galaxies, their visits to us can occur once in millions or even billions of years providing they can find Earth among billions of other stars and planets in the universe and if they were interested in visiting Earth millions or billions of years ago to arrive now. The range of our conjectures regarding the visits of aliens is: 1) none and never, which is the most probable, 2) rarely and by robotic starships with a time span between the visits from thousands of years to billions of years, 3) they possess a means for instant teleportation through space (which does not mean however that the time abyss between us and them regarding their search for potentially habitable planets by ordinary observations disappeared). Here, one may unleash all his imagination about green men and their presence among us.